\documentclass[preprintnumbers,showpacs,amsmath,amssymb,floatfix,prd,onecolumn,superscriptaddress,nofootinbib]{revtex4}
\usepackage{bbm}
\usepackage{graphicx}
\usepackage{epsfig}
\usepackage{bm}
\usepackage{amsfonts}
\usepackage{epstopdf}
\usepackage{multirow}

\begin{document}

\title{Bifurcation and Global Dynamical Behavior of the $f(T)$ Theory}

\author{Chao-Jun Feng}
\email{fengcj@shnu.edu.cn} \affiliation{Shanghai United Center for Astrophysics (SUCA), \\ Shanghai Normal University,
   100 Guilin Road, Shanghai 200234, P.R.China} 
   \affiliation{State Key Laboratory of Theoretical Physics, \\Institute of Theoretical Physics, Chinese Academy of Sciences, Beijing 100190, P.R.China}

\author{Xin-Zhou Li}
\email{kychz@shnu.edu.cn} \affiliation{Shanghai United Center for Astrophysics (SUCA),  \\ Shanghai Normal University,
   100 Guilin Road, Shanghai 200234, P.R.China}

\author{Li-Yan Liu}
\email{liuliyan081@126.com} \affiliation{Shanghai United Center for Astrophysics (SUCA),  \\ Shanghai Normal University,
    100 Guilin Road, Shanghai 200234, P.R.China}

\begin{abstract}
Usually, in order to investigate the evolution of a theory, one may find the critical points of the system and then perform perturbations around these critical points to see whether they are stable or not. This local method is very useful when the initial values of the dynamical variables are not far away from the critical points. Essentially, the nonlinear effects are totally neglected in such kind of approach. Therefore, one can not tell whether the dynamical system will evolute to the stable critical points or not when the initial values of the variables do not close enough to these critical points. Furthermore, when there are two or more stable critical points in the system, local analysis can not provide the informations that which one the system will finally evolute to.  In this paper, we have further developed the nullcline method to study the bifurcation phenomenon and global dynamical behaviour of the $f(T)$ theory.  We overcome the shortcoming of local analysis. And it is very clear to see the evolution of the system under any initial conditions.  
\end{abstract}

\pacs{98.80.-k, 95.36.+x, 02.30.Oz, 02.40.Xx }
\maketitle 
\flushbottom


\section{Introduction}\label{sec:intro}

As we known, the cosmological equations seem to be so complex due to the existence of a large number of variables, and it is almost impossible  to write down explicit solutions of nonlinear systems. Usually, one can use the technique of local dynamical analysis to determine the behavior of solutions near equilibrium or critical points \cite{Copeland:1997et}. The local method has been used to analyze many systems, such as the cosmological models, see Ref.~\cite{Hao:2003th,Li:2003ft,Li:2009zzc,Liu:2005wga,Ao:2010mg,Hao:2004uc,Liu:2010qq,Zhang:2010ym} .  In such kind of analysis, an equilibrium is said to be stable if nearby solutions stay nearby for future time. Thus, for  a given initial condition, the solution will or will not evolute to these stable critical points depending on the distance between the initial point and the critical point. Furthermore, one can often find that there are more than one stable critical points in a given system. By using the local analysis, it is easily to see that the system will finally evolute to one of these critical points as long as  the initial values are near this critical point. But for an arbitrary initial condition, local method can not be able to tell which stable point the system will evolute to. Therefore, only when the global behavior of dynamics system is fully understood, the fate of the universe have definitive decision. 

One of the most useful tools for analyzing nonlinear systems of differential equations are the nullclines, especially for planar systems. Also, there are usually some parameters in the systems. The behavior of the system will be different when these parameters take different values, and we call this kind of phenomena bifurcation. In \cite{Feng:2012wx}, we have developed a global method based on nullclines to analyze the global behavior of dynamical systems and also the bifurcation phenomena. In this paper, we will further develop this method and use it to study the bifurcation phenomenon and global dynamical behavior of the $f(T)$ theory. 

The $f(T)$ theory is a kind of modified gravity theory proposed by extending the action  of teleparallel gravity \cite{Hayashi:1979qx, Hehl:1976kj,Garecki:2010jj} in analogy to $f(R)$ theory. In the teleparallel gravity theory, one defines the so-called Weitzenb\"oak connection  on a curvature-free manifold in stead of the Levi-Civita connection in general relativity. As a result, the space-time has only torsion. In fact, this kind of description is equivalent to that of general relativity. It has been demonstrate that $f(T)$ theory can not only explain the present cosmic acceleration without dark energy\cite{Bengochea:2008gz,Bengochea:2010sg,Linder:2010py}, but also provide an alternative candidate to inflation\cite{Ferraro:2006jd, Ferraro:2008ey}. Observation constants on the $f(T)$ theory have been made in Ref.~\cite{Wu:2010mn,Wu:2010av,Zheng:2010am,Li:2010cg,Yang:2010ji}. New types of $f(T)$ theories have been proposed in Refs.\cite{Myrzakulov:2010vz,Myrzakulov:2010tc,Tsyba:2010ji,Yang:2010hw}. Background and perturbations analysis has been made in Refs.~\cite{Dent:2011zz,Chen:2010va,Bamba:2010wb}. Local analysis on $f(T)$ models with different variables and forms are performed in Refs.~\cite{Wu:2010xk,Zhang:2011qp}. For recent progress and reviews on $f(T)$ theory, see Refs.~\cite{Dong:2013rea,Bamba:2013fta,Huang:2013una,Myrzakulov:2012sp}.

This paper is organized as follows. In next section, we will briefly review of the $f(T)$ theory; in Sec.\ref{sec:null}  we shall perform the global analysis to the $f(T)$ model, in particular to the power law model and the logarithmic model, including the nullcline, bifurcation and phase portrait analysis. In Sec.\ref{sec:cos}, we will discuss the cosmological consequence of the $f(T)$ theory. In the final section, we will draw our conclusions and give some discussions.

\section{Briefly review of the $f(T)$ theory}\label{sec:ft}
The action of $f(T)$ theory reads
\begin{equation}\label{equ:action}
	S= \frac{1}{16\pi G} \int d^4x |e| \bigg[ T + f(T) \bigg]+ \int d^4x |e|\mathcal{L}_m \,,
\end{equation}
where $T$ is the torsion scalar, $|e|=\det(e^A_\mu) = \sqrt{-g}$ and $\mathcal{L}_m$ is the matter Lagrangian. Here $e^A_{\;\;\mu}(x)$ are the components of the vierbein vector field $\mathbf{e}_A$ in the coordinate basis $\mathbf{e}_A \equiv e^{\;\;\mu}_A\partial_\mu$. Note that in the teleparallel gravity, the dynamical variable is the vierbein field $\mathbf{e}_A(x^\mu)$.  Consider the following metric
\begin{equation}\label{equ:metric}
	ds^2= g_{\mu\nu}dx^\mu dx^\nu = \eta_{ab}\theta^A\theta^B \,,
\end{equation}
 where $g_{\mu\nu}$ being the metric of space-time, the Minkowski's metric $\eta_{AB}= \text{diag}(-1,1,1,1)$, the tetrads $\theta^A = e^A_{\;\;\mu} dx^\mu$ and their inverse $e^{\;\;\mu}_A$ the tetrads basis $dx^\mu = e_A^{\;\;\mu} \theta^A $. Then, the basis satisfy the relations
\begin{equation}\label{equ:relate}
	e^A_{\;\;\mu} e^{\;\;\nu}_A = \delta^\nu_\mu \,, \quad e^A_{\;\;\mu} e^{\;\;\mu}_B = \delta^A_B \,,
\end{equation}
where $A, B$ are indices running over $0,1,2,3$ for the tangent space of the manifold and $\mu, \nu$ are coordinate indices on the manifold, also running over $0,1,2,3$.  Thus, the vierbein field is related with the space-time metric by
\begin{equation}\label{equ:relate2}
	g_{\mu\nu} = \eta_{AB}e^A_{\;\;\mu} e^B_{\;\;\nu} \,,
\end{equation}
and the root of the metric determinant is given by $|e|=\sqrt{-g} = \det(e^A_{\;\;\mu})$.

In the teleparallel gravity theory, we use the standard Weitzenb\"ok's connection defined as
\begin{equation}\label{equ:weicon}
	\Gamma^\alpha_{\mu\nu} = e_A^{\;\;\alpha} \partial_\nu e^A_{\;\;\mu} = -e^A_{\;\;\mu}\partial_\nu e^{\;\;\alpha}_A \,.
\end{equation}
And the covariant derivative $D_\mu$ satisfies the equation 
\begin{equation}\label{equ:cd}
	D_\mu e^A_\nu = \partial_\mu e^A_{\;\;\nu} - \Gamma^\alpha_{\nu\mu} e_{\;\;\alpha}^A = 0 \,.
\end{equation}
Then the components of the torsion and contortion tensors are given by
\begin{eqnarray}
	T^\alpha_{\;\;\mu\nu} &=& \Gamma^\alpha_{\nu\mu} - \Gamma^\alpha_{\mu\nu} = e^{\;\;\alpha}_A(\partial_\mu e^A_{\;\;\nu} - \partial_\nu e^A_{\;\;\mu}) \,,\\
	K^{\mu\nu}_{\;\;\;\;\alpha} &=& -\frac{1}{2}\bigg(T^{\mu\nu}_{\;\;\;\;\alpha} - T^{\nu\mu}_{\;\;\;\;\alpha} - T^{\;\;\mu\nu}_\alpha\bigg)\,.
\end{eqnarray}
By introducing another tensor 
\begin{equation}\label{equ:stensor}
	S_\alpha^{\;\;\mu\nu} = \frac{1}{2}\bigg(K^{\mu\nu}_{\;\;\;\;\alpha} +\delta^\mu_\alpha T^{\beta\nu}_{\;\;\;\;\beta} - \delta^\nu_\alpha T^{\beta\mu}_{\;\;\;\;\beta} \bigg) \,,
\end{equation}
we can define the torsion scalar as 
\begin{equation}\label{equ:torsion scalar}
	T \equiv T^\alpha_{\;\;\mu\nu} S_\alpha^{\;\;\mu\nu} \,.
\end{equation}

After applying the action principle with respect to the vierbein field, we obtain the equation of motion as
\begin{equation}\label{equ:eom}
	e_A^{\;\;\beta}S_\beta^{\;\;\mu\alpha} (\partial_\alpha T )f_{TT} 
	+\bigg[  |e|^{-1} \partial_\alpha(|e|e^{\;\;\sigma}_A S_\sigma^{\;\;\mu\alpha}) 
		  + e_A^{\;\;\beta}T^\sigma_{\;\;\nu\beta}S_\sigma^{\;\;\mu\nu}\bigg] (1+f_T) 
		  +\frac{1}{4}e_A^{\;\;\mu} f 
		  = 4\pi G e_A^{\;\;\beta}T^\mu_\beta \,,
\end{equation}
where the subscript $T$ denotes derivatives with respect to $T$.
For simplicity, we assume a flat Friedmann-Robertson-Walker metric,
\begin{equation}\label{equ:FRW}
	ds^2 = -dt^2 + a(t)^2 (dx^i)^2 \,,
\end{equation}
 with scale factor $a(t)$. So we have $e^A_{\;\;\mu} = \text{diag}(1,a,a,a)$ and the torsion scalar $T = -6H^2$, where $H=\dot a/a$ is the Hubble parameter. The modified Friedmann equations then read
 \begin{eqnarray}
	H^2 &=& \frac{1}{3}(\rho_m + \rho_r ) - \frac{f}{6} -2 H^2 f_T  \,,\\
	(H^2)' &=& \frac{2p + 6H^2 + f + 12H^2f_T}{24H^2f_{TT}-2-2f_T} \,,
\end{eqnarray}
where the prime denotes the derivative with respect to $\ln a$ and $p=\rho_r/3$. Here and after, we use the units $8\pi G = 1$.  The energy conservation equations for the radiation and matter are as follows
\begin{eqnarray}
	\rho_r' &+& 4\rho_r = 0 \,,\\
	\rho_m' &+& 3\rho_m = 0 \,.
\end{eqnarray}
One can also define an effective dark energy with the energy density 
\begin{equation}\label{equ:eff}
	\rho_e = \frac{1}{2}(-f+2Tf_T) \,,
\end{equation}
and equation of state 
\begin{equation}\label{equ:eos}
	w_e = -1 + \frac{1}{3}\frac{T'}{T} \frac{f_T+2Tf_{TT}}{f/T-2f_T} \,,
\end{equation}
which can be derived from the conservation law.

\section{Global analysis: nullcline, bifurcation and phase portrait}\label{sec:null}
In this section, we shall use the qualitative technique of nullcline for analyzing the global behavior of nonlinear system and study the bifurcation phenomena in the dynamical systems. First we introduce the following dimensionless variables 
\begin{equation}\label{equ:dimless}
	x = \frac{\rho_e}{3H^2}\,,\quad y =\Omega_m=\frac{\rho_m}{3H^2}\,,\quad z = \Omega_r = \frac{\rho_r}{3H^2} \,
\end{equation}
and then the dynamical equations of the nonlinear system could be rewritten as follows
\begin{eqnarray}
	x' &=& \left(f_T - \frac{f}{T} - 2Tf_{TT} \right)\frac{T'}{T} \,, \label{equ:dy1}\\
	y' &=& -y\left(3+\frac{T'}{T}\right)\,, \label{equ:dy2}
\end{eqnarray}
where we have used the Friedmann equation $x+y+z=1$ and here
\begin{equation}\label{equ:TT}
	\frac{T'}{T} = \frac{(H^2)'}{H^2} = -\frac{4-4x-y}{2Tf_{TT} + f_T +1} \,.
\end{equation}
Once the function $f(T)$ is specified, it is possible to express $T$ as function of $x$ by using the definition of $x$, then one can analysis the dynamical system (\ref{equ:dy1}) and (\ref{equ:dy2}) more detail. In this paper, we will focus on two of the most popular forms of the function $f$: one is the power law model, the other is the logarithmic model.

\subsection{Power law model}
First we will consider the pow law model with the function $f$ as follows \cite{Linder:2010py}
\begin{equation}\label{equ:power}
	f(T) = \alpha (-T)^n \,,
\end{equation}
where $\alpha$ and $n$ are dimensionless parameters. In the case of $n=0$, it reduces to the $\Lambda$CDM model.
In this model, the energy density (\ref{equ:eff}) and $x$ are given by
\begin{eqnarray}
	\rho_{e} &=&  \alpha\left(  n-\frac{1}{2}\right)(-T)^n \,,\\
	x &=&  \alpha\left(  2n-1\right)(-T)^{n-1}\,,
\end{eqnarray}
and the evolution of $T$ can be simplified as
\begin{equation}
	\frac{T'}{T} = -\frac{4-4x-y}{-n(2n-1)\alpha(-T)^{n-1} +1} 
	= \frac{4-4x-y}{nx-1} \,.
\end{equation}
Then, the dynamical equations (\ref{equ:dy1}) and (\ref{equ:dy2}) are
\begin{eqnarray}
	x' &=& \frac{(n-1)(4-4x-y)x}{nx-1} \,, \label{equ:dy1p}\\
	y' &=& -\frac{[1+(3n-4)x-y]y}{nx-1}\,. \label{equ:dy2p}
\end{eqnarray}
It is obvious that when $n=1$, $x=x_0$ is a constant, and then 
\begin{equation}
	y' = y-\frac{y^2}{1-x_0}\,,
\end{equation}
which has a solution
\begin{equation}
	y = (1-x_0) \left[ \left(\frac{1-x_0}{y_0} - 1\right)\frac{1}{a}+ 1\right]^{-1}\,,
\end{equation}
 where $x_0, y_0$ denote the  values of $x, y$ at $\ln a = 0$ or $a=1$. Therefore, the system will evolute to the point $(x_0, 1-x_0)$ in the future when $\ln a\rightarrow \infty$. Actually, the case $n=1$ is just equivalent to rescaling the Newton's constant $G$. Therefore, we will focus on the case of $n\neq 1$ in the following.

For the system (\ref{equ:dy1p}) and (\ref{equ:dy2p}), the $x$-nullclines are the set of points determined by
\begin{equation}\label{equ:xnull}
	x=0\,,\quad 4x+y=4 \,,
\end{equation}
while the $y$-nullclines are determined by
\begin{equation}\label{equ:ynull}
	y=0\,,\quad (3n-4)x - y = -1 \,.
\end{equation}
The intersections of $x$- and $y$- nullclines yield the equilibrium points 
\begin{equation}\label{equ:ep}
	(0\,,\; 0 )\,,\quad (0\,,\; 1)\,,\quad (1\,,\;0)  \,.
\end{equation}
It should be noticed that the $x$- and $y$- nullclines also meet at $(1/n, 4-4/n)$, but it is not a equilibrium point of the system, because at this point $x'\rightarrow 4(1-n)/n^2 $ and $y'\rightarrow 4(3n-4)(1-n)/n^2 $, which can not be vanished simultaneously when $n\neq 1$.  However, $(1/n, 4-4/n)$ is still a `critical' point since $x'$ and $y'$ are divergent on the line $x=1/n$ except $(1/n, 4-4/n)$. The line $x=1/n$ played a crucial geometric role, which determined the direction of the trajectories as $\ln a\rightarrow \pm \infty$. A trajectory starting on this line stays on it forever when $y>4-4/n$ and $n>1$ or when $y<4-4/n$ and $n<1$. In fact, these semi-lines are globally attracting. Furthermore, this system has a singularity at $x=1/n$, which does not affect the continuity of the solution curves because the limit of $dy/dx$ does exist. The only effect of the singularity is  the direction of the vector field or the flow of solutions will change when they cross the line $x=1/n$. One can blow up the singularity at $x=1/n$ by introducing a new variable $s$ via the rule $d\ln a /ds = (nx-1)^2$. And then, the system becomes
\begin{eqnarray}
	\dot x &=& (n-1)(4-4x-y)(nx-1)x \,, \label{equ:dy1pbu}\\
	\dot y &=& -[1+(3n-4)x-y](nx-1)y \,, \label{equ:dy2pbu} 
\end{eqnarray} 
where the dot indicates differentiation with respect to $s$. The solution curves of the new system (\ref{equ:dy1pbu}) and (\ref{equ:dy2pbu}) including their directions remain the same as system (\ref{equ:dy1p}) and (\ref{equ:dy2p}), but they are parameterized differently.  Further more, for the new system we have
\begin{eqnarray}
	\dot y &=& -(y-1)y \,, \qquad \qquad \; \qquad \text{on} \quad x=0 \,, \\
	\dot y &= &-3y\left(1-n+\frac{n}{4}y\right)^2 \,, \qquad \text{on} \quad x=(4-y)/4 \,.
\end{eqnarray}
and also
\begin{eqnarray}
	\dot x &=& -4n(n-1) (x-1)\left(x-\frac{1}{n}\right)x \,, \quad \text{on}\quad y=0 \,, \\
	\dot x &=& -3(n-1)(nx-1)^2 x \,,\qquad \qquad \quad \text{on} \quad y=1+(3n-4)x \,.
\end{eqnarray}
Obviously, the dynamical behavior of the system is depending on the parameter $n$, which means different values of $n$ will determine different evolution behaviors of the system. This is the so-called bifurcation phenomena. Although the system we studied here is planar and one can see the bifurcation phenomena clearly after plotting the phase portrait with different values of $n$, we can still catch the main properties of the bifurcations on different nuclines. 

\begin{figure}[h]
\begin{center}
\includegraphics[width=0.5\textwidth,angle=0]{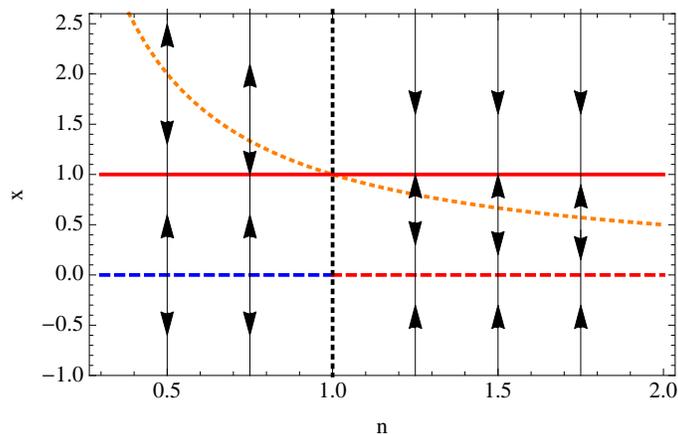}
\caption{\label{fig:bf1}The bifurcation diagram on the nullcline $y=0$ for the power law model.  } 
\end{center}
\end{figure}

On the $x$-nullcline $x=0$, the evolution of $y$ is not depending on $n$, while on the $y$-nullcline $y=0$, the evolution of $x$ is indeed depending on $n$ and then the bifurcation happens when $n$ changes from $n<1$ to $n>1$ or the inverse.  We have plot the bifurcation diagram in Fig.\ref{fig:bf1}, in which the arrow denotes the time direction or the flow of the solutions. Equivalently, it also denotes the vector direction, i.e.$(\dot x, \dot y)$, of the system. 

Actually, the nullclines could tell us  the behaviors of the system not only near the critical points but also at the points far away. As we mentioned before, the evolution of $y$ is independent of $n$ on $x=0$, therefore, the vector field is tangent to the $y$-axis, and the solutions tend away from $(0,0)$ and tend to $(0,1)$ along the $y$-axis.  However,  the vector field  is not tangent to the second $x$-nullcline, i.e. $x=(4-y)/4$. One can see that the  vector on this nullcline points south when $y>0$ and north when $y<0$.  This will be hold in the case of $n<1$.  On the other side, the evolution of $x$ on the $y$-nullclines is depending on $n$. We plot the phase portraits of the system with the parameter $n=2, 1/2 , 0 ,-2$ in Fig.\ref{fig:fig1}, in which one can clearly see the bifurcation phenomenon and global behavior of the $f(T)$ theory with power law form. We also give some discussions in each cases as follows.

\subsubsection{Case: $n>1$}
In this case, the solutions tend to $(1,0)$ and $(0,0)$  and tend away from $x=1/n$ along the line $y=0$. The vector field on the nullcline $y=1+(3n-4)x$ points east when $x<0$ and west when $x>0$, see Fig.\ref{fig:fig1}~(left top). From this figure, one can clearly see that there are two sinks at $(0,1), (1,0)$ and one saddle point at $(0,0)$.

On the other side, from Eqs.~(\ref{equ:dy1p}) and (\ref{equ:dy2p}), the curves in the phase portrait always satisfies the following equation
\begin{equation}\label{equ:phase}
	\frac{dy}{dx} = \left(\frac{dx}{dy}\right)^{-1}= \frac{-[1+(3n-4)x-y]y}{(n-1)(4-4x-y)x} \,,
\end{equation}
and
\begin{eqnarray}
	\frac{d^2y}{dx^2}&=&\frac{ny (1-x-y)}{(n-1)^2 x^2 (4-4 x-y)^3}  (y-y_+)(y-y_-) \,,\\
	\frac{d^2x}{dy^2}&=&\frac{n(n-1) x (1-x-y)}{y^2 (1+(3 n-4) x-y)^3}  (y-y_+)(y-y_-) \,,
\end{eqnarray}
where
\begin{equation}
	y_\pm = 2 \left[2 (1-x) \pm \sqrt{3n} \left|x-\frac{1}{n}\right|\right]\,.
\end{equation}
In the limit of  $x\rightarrow 1/n + \delta $, we have
\begin{equation}\label{equ:phase2}
	\frac{dy}{dx} = \left(\frac{dx}{dy}\right)^{-1} =-\frac{n}{n-1} \left[ 1-\frac{3n\delta}{y-(4-4/n -4 \delta )}\right] \frac{y}{1+n\delta}  \,.
\end{equation}
and
\begin{equation}
	y_\pm = 4-\frac{4}{n} - 4\delta \pm 2\sqrt{3n} |\delta| \,.
\end{equation}
Furthermore, when $y= 4-4/n -4 \delta $, $dy/dx \rightarrow \infty$ or $dx/dy \rightarrow 0$, we get
\begin{equation}
	\frac{d^2x}{dy^2}=\frac{(n-1)}{12 n\delta }\frac{(1+n\delta)}{(n-1-n\delta) } 
	\approx \frac{1}{12 n\delta }\,,
\end{equation}
with $|\delta| \ll |1/n|$. Thus,  the curves of solutions that passed the point $(1/n+\delta, 4-4/n -4 \delta)$  will be concave right $(\delta >0)$ or left ($\delta <0$), which means they will never hit the line $x=1/n$, and finally flow to the point $(1,0)$ when $\delta >0$ and to the point $(0,1)$ when $\delta<0$.  Also, we have $dy/dx = 0$ when $y\rightarrow4-4/n +(3n-4) \delta $, and
\begin{equation}
	\frac{d^2y}{dx^2}=\frac{(3n-4)}{3n(n-1)\delta }\frac{(4n-4+(3n-4)n\delta) }{(1+n\delta) }
	\approx \frac{4}{3n \delta}(3n-4)\,.
\end{equation}
Thus, the curves of solutions that passed the point $(1/n+\delta, 4-4/n +(3n-4) \delta)$ will be concave up when $\delta>0$ and $n>4/3$ (or $\delta<0$ and $n<4/3$), which means they  will  hit the line $x=1/n$ and never flow to the point $(1,0)$ (or flow to the point $(0,1)$), while the curves will be concave down when $\delta >0$ and $n<4/3$ (or $\delta<0$ and $n>4/3$), which means  they will flow to the point $(1,0)$ (or flow to the point $(0,1)$ or $(0,0)$), see Fig.\ref{fig:fig1}~(left top).  By taking  the limit of $\delta \rightarrow 0$, we can get the following results: for the solutions with the initial condition $x_i<1/n$ will finally flow to the sink point $(0,1)$ or the saddle point $(0,0)$, while for the solutions  with initial conditions $(x_i>1/n, y_i<4-4/n)$ will finally flow to the sink point $(1,0)$, and the solutions with $(x_i>1/n, y_i\gg 4-4/n)$ will finally hit the line $x=1/n$ instead of the three equilibrium points.

\subsubsection{Case: $0<n<1$}
In this case, the vertical line $x=1/n$ will move to the right side of $x=1$. Then, the solutions tend to $(1,0)$ and tend away from $(0,0)$ and $x=1/n$ along the line $y=0$. The vector field on the nullcline $y=1+(3n-4)x$ points west when $x<0$ and east when $x>0$, see Fig.\ref{fig:fig1}~(right top).

With the initial condition $0<x_i<1/n$, the solutions will eventually flow to the sink point $(1,0)$, while with the initial condition $x_i<0$ or $x_i>1/n$, the solutions will not flow to any equilibrium points. Considering the  physical conditions $y, z>0$ and the constraint $x+y+z=1$, we know that $x<1$, but $x$ could be negative in the early universe, because it is only an effective energy density of dark energy. So, one should be very careful to choice the condition $x_i>0$ in order to get the final state at point $(1,0)$ with dark energy dominated.  Furthermore, here we only consider the classical evolution, so even if $x_i>0$, it could evolute to $x<0$ by the quantum effect. 

\subsubsection{Case: $n= 0$}
This is a limit case of $ 0<n<1$, and the vertical line $x=1/n$ disappears or it moves to infinity when $n\ll 1$. Then once $x_i>0$, the system will eventually move to the sink point $(1,0)$, see Fig.\ref{fig:fig1}~(left bottom). 

\subsubsection{Case: $n<0$}
\begin{figure}[h]
\begin{center}
\includegraphics[width=0.4\textwidth,angle=0]{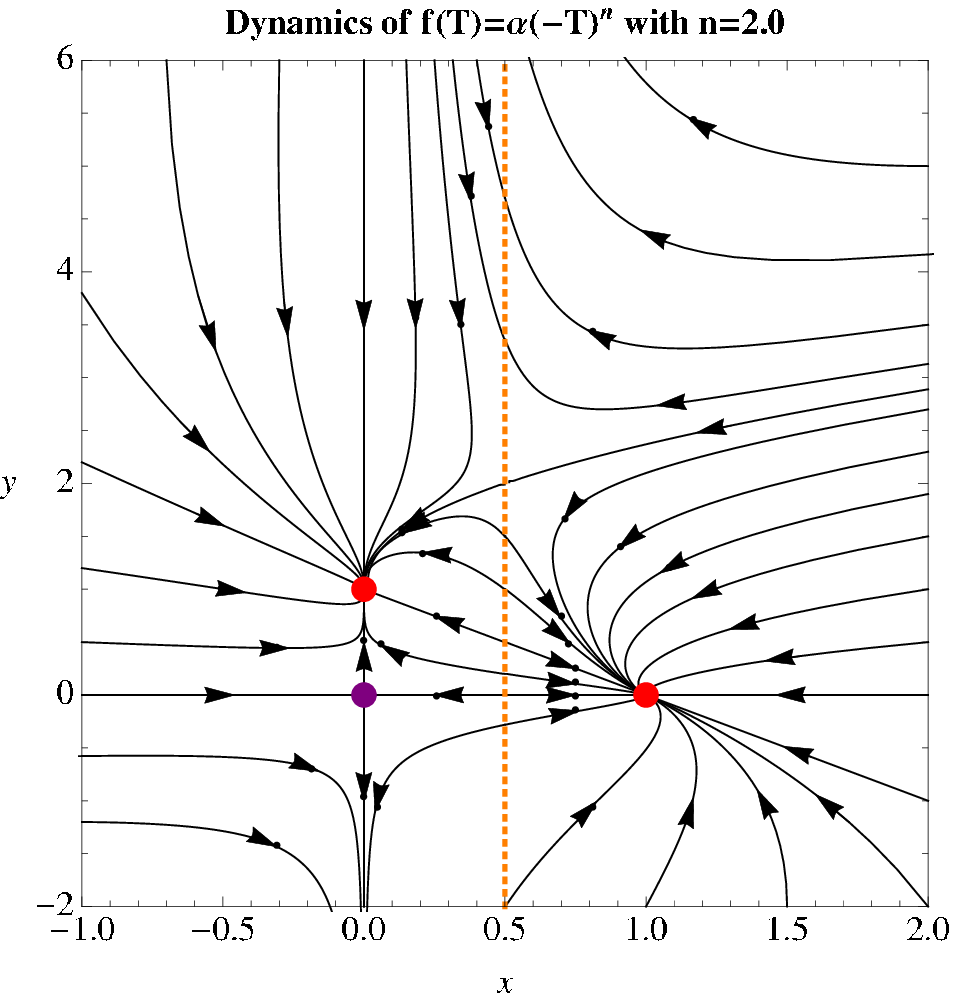}
\qquad
\includegraphics[width=0.4\textwidth,angle=0]{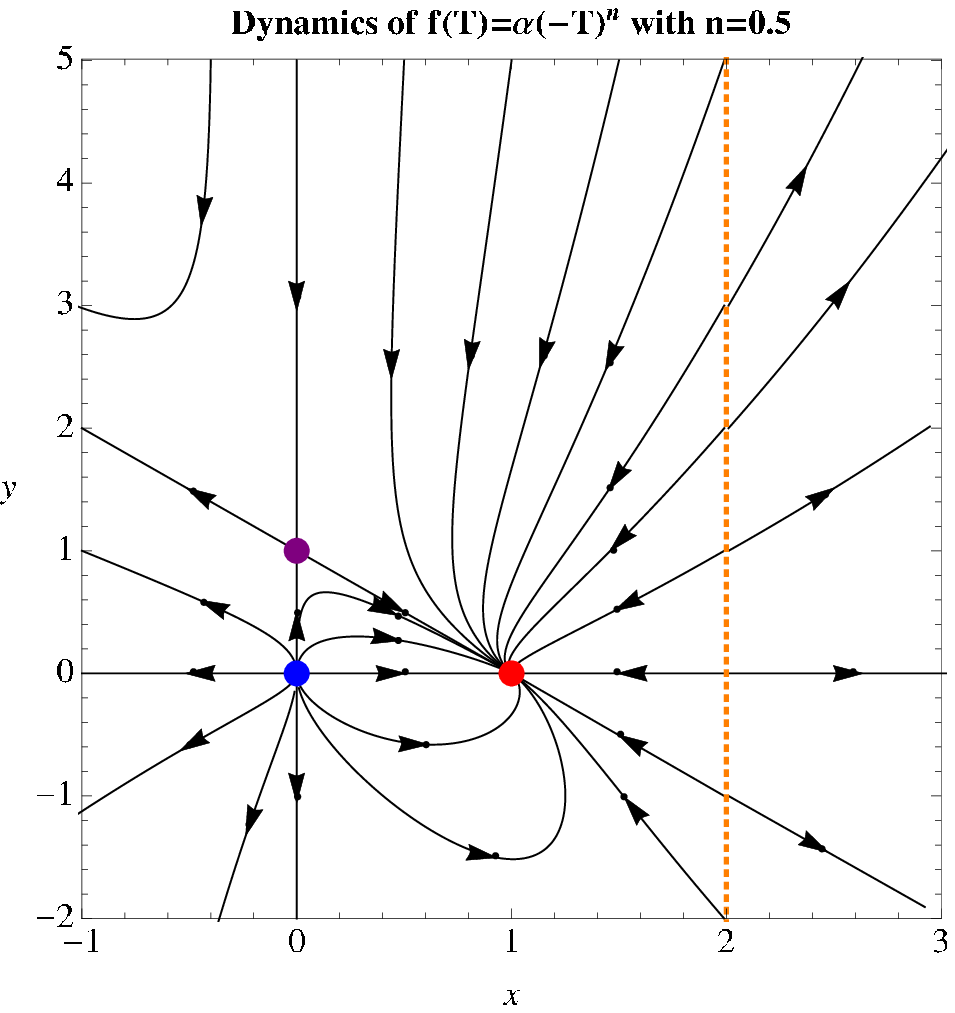}
\qquad
\includegraphics[width=0.4\textwidth,angle=0]{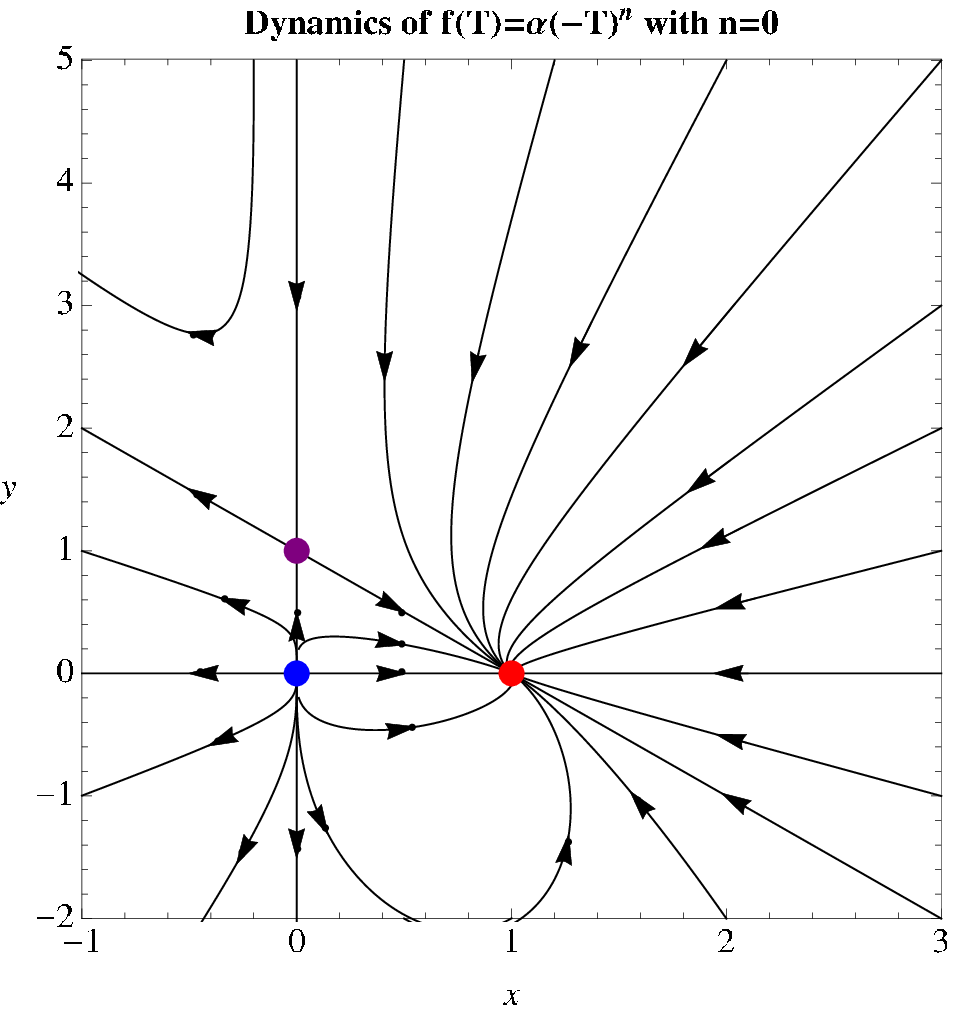}
\qquad
\includegraphics[width=0.4\textwidth,angle=0]{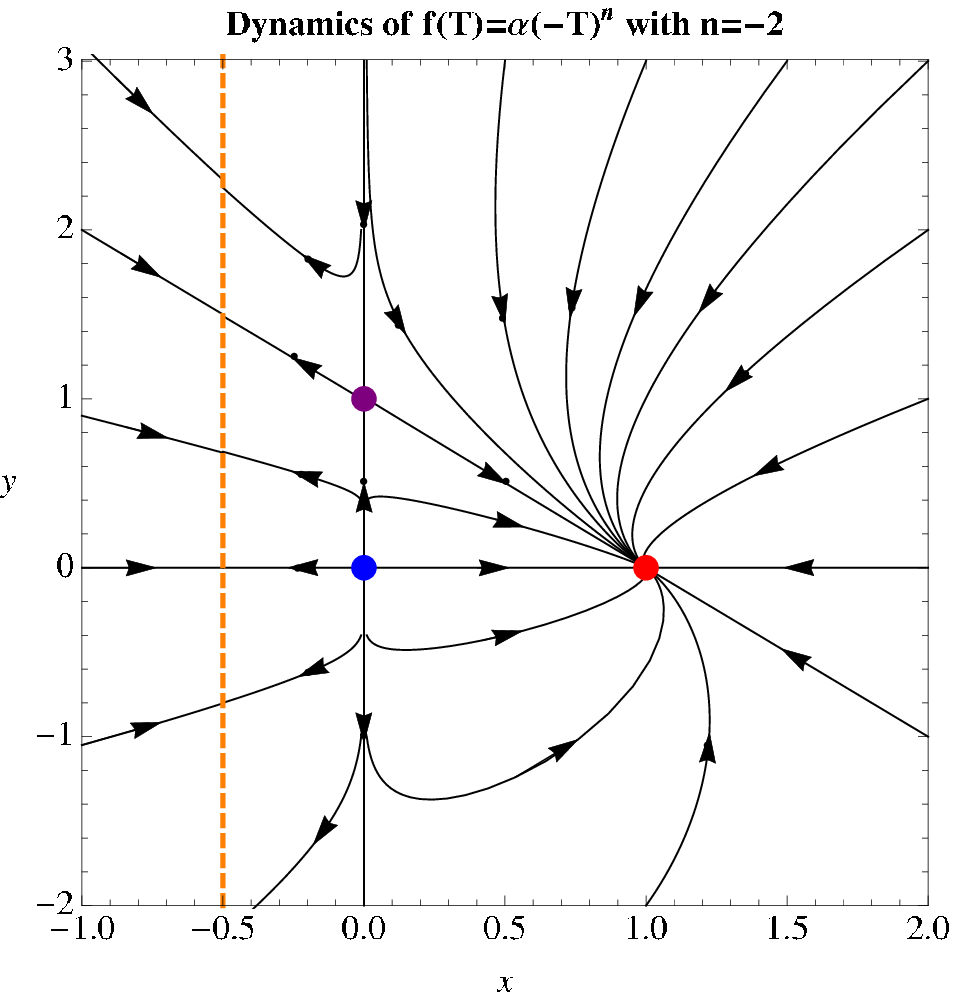}
\caption{\label{fig:fig1}The phase portraits of the power law model with $n=2.0$(left top), $0.5$(right top), $0$(left bottom) and $-2.0$(right bottom), respectivly. The equilibrium  point at $(1,0)$ is always a sink point (red), that at $(0,0)$ is a source ($n<1$) point (blue) or a  saddle ($n>1$) point (purple), and that at $(0,1)$ is a sink ($n>1$) point (red) or a saddle ($n<1$) point (purple).  The vertical dotted line (orange) denotes $x=1/n$. The arrow denotes the direction of the vector field $(\dot x, \dot y)$ or equivalently the flow of the solutions. } 
\end{center}
\end{figure}

In the case of $n<0$, the vertical line $x=1/n$ will move to the left side of $x=0$. Then, the solutions tend to $(1,0)$ and $x=1/n$  and tend away from $(0,0)$ along the line $y=0$. Then once $x_i>0$, the system will eventually move to the sink point $(1,0)$, see Fig.\ref{fig:fig1}~(right bottom).

In summary, we have shown that in  the case of $n\neq 0$, there is always a vertical line $x=1/n$ that changes the direction of the vector field or the flow of the solutions. If   a real physical system requires the direction of flow of the solutions should not be changed until they flow to the equilibrium points, then there are only two cases satisfy this condition: one is $n=0$ corresponding to $\Lambda$CMD model, the other is $|n|\ll 1$ corresponding to a variable $f(T)$ model with small running of the equation of state \cite{Linder:2010py}.

\subsection{Logarithmic model}
In this section, we will consider the logarithmic model with the function $f$ as follows \cite{Myrzakulov:2010vz}
\begin{equation}\label{equ:log}
	f(T) = \alpha T^\beta \ln T \,, 
\end{equation}
where $\alpha$ and $\beta$ are dimensionless parameters. When $\alpha=0$, it reduces to the $\Lambda$CDM model, and then we will consider the case $\alpha\neq0$ in the following.   In this model, the energy density (\ref{equ:eff}) and $x$ are given by 
\begin{eqnarray}
	\rho_{e} &=&  \alpha T^\beta \bigg[ \left(\beta -\frac{1}{2}\right)\ln T+1 \bigg] \,,\\
	x &=&  -\alpha T^{\beta-1} \bigg[ \left(2\beta -1\right)\ln T+2 \bigg] \,,
\end{eqnarray}
and the evolution of $T$ can be simplified as
\begin{equation}\label{equ:TT}
	\frac{T'}{T} = -\frac{4-4x-y}{\alpha T^{\beta -1} [\beta(2\beta-1)\ln T + 4\beta -1] +1} \,,
\end{equation}
In the cases of $\beta=1/2$ and $\beta=1$ which we are most interested, we can express the above equation in terms of $x,y$ to get the plane systems. So, in the following, we will focus on these two cases.

\subsubsection{Case: $\beta =1/2$}
\begin{figure}[h]
\begin{center}
\includegraphics[width=0.4\textwidth,angle=0]{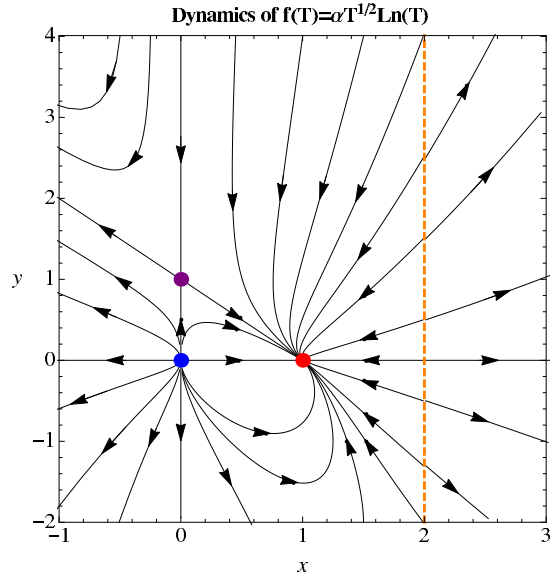}
\caption{\label{fig:figlog1} The phase portrait of the logarithmic model with $\beta=1/2$. The equilibrium  point at $(1,0)$ is  a sink point (red), that at $(0,0)$ is a source point (blue), and that at $(0,1)$ is a saddle point (purple).  The vertical dotted line (orange) denotes $x=2$. The arrow denotes the direction of the vector field $(\dot x, \dot y)$ or equivalently the flow of the solutions. } 
\end{center}
\end{figure}

In this case, $x=-2\alpha T^{-1/2}$, and Eq.(\ref{equ:TT}) is simplified to
\begin{equation}
	\frac{T'}{T} = -\frac{4-4x-y}{\alpha T^{-1/2}  +1} = 2 \frac{4-4x-y}{x-2}\,,
\end{equation}
Then, the dynamical equations (\ref{equ:dy1}) and (\ref{equ:dy2}) are given by
\begin{eqnarray}
	\dot x &=&  - (4-4x-y)x(x-2)\,, \label{equ:dy1log1}\\
	\dot y &=& -(2-5x-2y)y(x-2) \,. \label{equ:dy2log1}
\end{eqnarray}
Obviously, the system is independent of parameters, so there is no bifurcation phenomenon in the case of $\beta=1/2$. Here, we have also blown up  the singularity at $x=2$ by introducing a new variable $s$ via the rule $d\ln a/ds = (x-2)^2$, and the dot indicates differentiation with respect to $s$. For the original system, the $x$-nullclines are given by
\begin{equation}
	x = 0\,,\quad 4x+y=4 \,,
\end{equation}
while  the $y$-nullclines are given by
\begin{equation}
	y=0\,,\quad 5x+2y=2 \,.
\end{equation}
Thus, the equilibrium points determined by the intersections of $x$- and $y$-nullclines are
\begin{equation}
	(0\,,\; 0 )\,,\quad (0\,,\; 1)\,,\quad (1\,,\;0)  \,.
\end{equation}
It should be noticed that the $x$- and $y$- nullclines also meet at $(2, -4)$, but it is not a equilibrium point of the system, because at this point $x'\rightarrow 8 $ and $y'\rightarrow -20 $.  For the system (\ref{equ:dy1log1}) and (\ref{equ:dy2log1}), we have
\begin{eqnarray}
	\dot y &=& -4(y-1)y \,, \qquad  \, \qquad \text{on} \quad x=0 \,, \\
	\dot y &= &-\frac{3}{16}y(y+4)^2 \,, \quad \qquad \text{on} \quad x=(4-y)/4 \,.
\end{eqnarray}
So the solutions tend away from $(0,0)$ and tend to $(0,1)$ along the $y$-axis. On the other side, we also have
\begin{eqnarray}
	\dot x &=& 4x(x-1)(x-2) \,, \quad \quad \text{on}\quad y=0 \,, \\
	\dot x &=& \frac{3}{2}x(x-2)^2 \,, \;\,\qquad \qquad \text{on} \quad y=1-\frac{5}{2}x \,.
\end{eqnarray}
So the solutions tend to $(1,0)$ and tend away from $(0,0)$ and $x = 2$ along the $x$-axis. Actually, this system (\ref{equ:dy1log1}) and (\ref{equ:dy2log1}) are just the same as that in the power law model with $n=0.5$ up to some constants. For completeness, we also plot the phase portrait for the model in Fig.\ref{fig:figlog1}.  

\subsubsection{Case: $\beta =1$}
In this case, $x=-\alpha(\ln T +2)$, and Eq.(\ref{equ:TT}) is simplified to
\begin{equation}
	\frac{T'}{T} = -\frac{4-4x-y}{\alpha(\ln T +3)  +1} =  \frac{4-4x-y}{x-(1+\alpha)}\,.
\end{equation}
Thus, the dynamical equations (\ref{equ:dy1}) and (\ref{equ:dy2}) become
\begin{eqnarray}
	\dot x &=& -\alpha(4-4x-y)(x-1-\alpha)\,, \label{equ:dy1log2}\\
	\dot y &=& -y(1-3\alpha - x - y)(x-1-\alpha)\,. \label{equ:dy2log2}
\end{eqnarray}
where we have also blown up  the singularity at $x=2$ by introducing a new variable $s$ via the rule $d\ln a/ds = (x-1-\alpha)^2$, and the dot indicates differentiation with respect to $s$. For the original system, the $x$-nullclines are given by
\begin{equation}
	 4x+y=4 \,,
\end{equation}
while  the $y$-nullclines are given by
\begin{equation}
	y=0\,,\quad x+y=1-3\alpha \,.
\end{equation}
Thus, the only equilibrium point determined by the intersections of $x$- and $y$-nullclines is $(1,0)$. It should be noticed that the $x$- and $y$- nullclines also meet at $(1+\alpha, -4\alpha)$, but it is not a equilibrium point of the system, because at this point $x'\rightarrow 4\alpha \neq 0$ and $y'\rightarrow -4\alpha \neq 0$. For the system (\ref{equ:dy1log2}) and (\ref{equ:dy2log2}), we have
\begin{equation}
	\dot y = -\frac{3}{16}y(y+4\alpha)^2 \,, \quad \qquad \text{on} \quad x=(4-y)/4
\end{equation}
and
\begin{eqnarray}
	\dot x &=&  4\alpha(x-1)(x-1-\alpha) \,, \quad \quad \text{on}\quad y=0 \,, \\
	\dot x &=&3\alpha(x-1-\alpha)^2 \,, \;\;\qquad \qquad \text{on} \quad y=1-x-3\alpha \,.
\end{eqnarray}
Obviously, the dynamical behavior of the system is depending on the parameter $\alpha$, and there is indeed a  bifurcation phenomenon in the case of $\beta=1$.  Again, we can still catch the main properties of the bifurcations on different nullclines. 

\begin{figure}[h]
\begin{center}
\includegraphics[width=0.5\textwidth,angle=0]{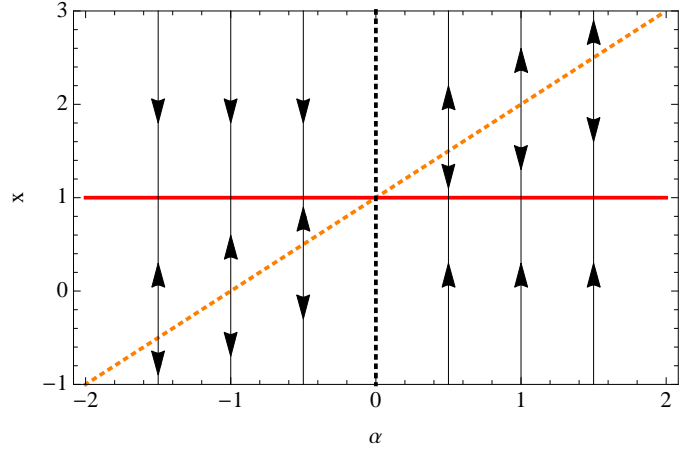}
\caption{\label{fig:bf2}  The bifurcation diagram on the nullcline $y=0$ for the logarithmic model with $\beta=1$.} 
\end{center}
\end{figure}

\begin{figure}[h]
\begin{center}
\includegraphics[width=0.4\textwidth,angle=0]{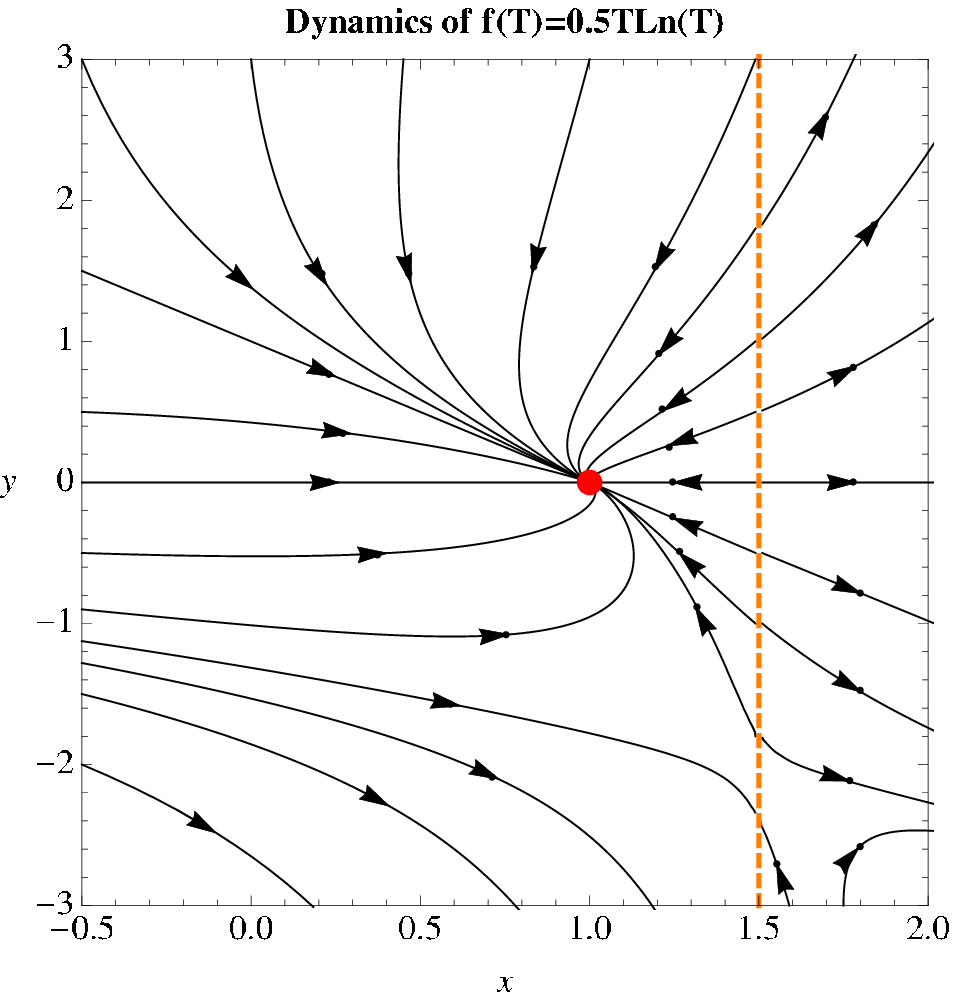}
\qquad
\includegraphics[width=0.4\textwidth,angle=0]{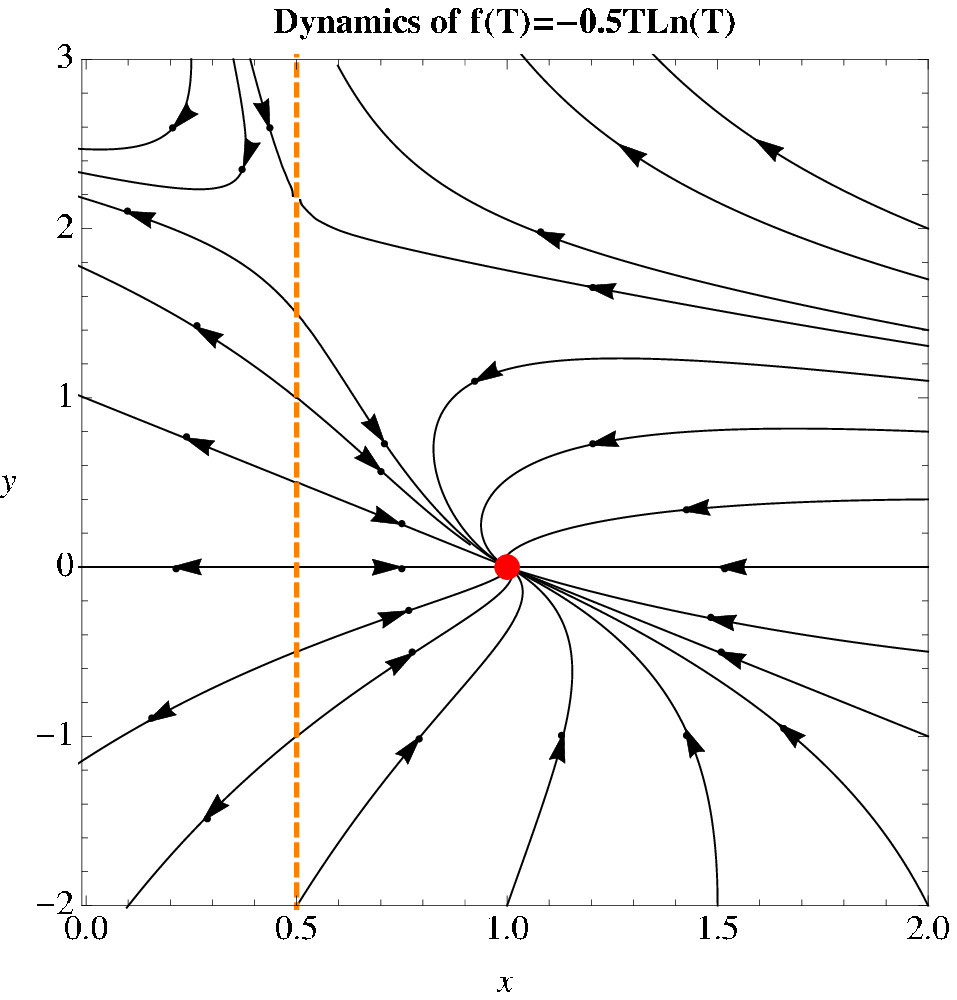}
\caption{\label{fig:fig2} The phase portraits of the logarithmic model with $\alpha=0.5$(left) and $-0.5$(right), respectivly. The equilibrium  point at $(1,0)$ is always a sink point (red). The vertical dotted line (orange) denotes $x=1+\alpha$. The arrow denotes the direction of the vector field $(\dot x, \dot y)$ or equivalently the flow of the solutions.  } 
\end{center}
\end{figure}

On the $y$-nullcline $y=0$, the evolution of $x$ is  depending on $\alpha$ and then the bifurcation happens when $\alpha$ changes from positive to negative or the inverse.  We have plot the bifurcation diagram in Fig.\ref{fig:bf2}, in which the arrow denotes the time direction or the flow of the solutions. Equivalently, it also denotes the vector direction, i.e.$(\dot x, \dot y)$, of the system. We Also plot the phase portraits of the system with the parameter $\alpha=\pm 1/2$ in Fig.\ref{fig:fig2}, in which one can clearly see the bifurcation phenomenon and global behavior of the $f(T)$ theory with logarithmic form.  Clearly, in the case of $\alpha>0$ ($\alpha<0$), the system will eventually evolute to the sate $(0,1)$ as long as the initial values of $x< 1+\alpha$ ($x>1+\alpha$), otherwise, it will never flow to any fix points. 

\section{Cosmological consequence of $f(T)$ theory}\label{sec:cos}

As we have seen in the previous section, the critical point $(1,0)$ is a sink in all the cases, then the cosmological system will eventually evolute to this sate as long as some initial conditions discussed before are satisfied. In fact, for an arbitrary function $f(T)$, $(1,0)$ is always a critical point, see Eqs.(\ref{equ:dy1}), (\ref{equ:dy2}) and (\ref{equ:TT}). This critical point corresponds to the state that the effective dark energy component dominates the universe. If the decay of the matter and radiation is so fast that the gravity has not become to general relativity, the universe will definitely go to this state, or the cosmological system will finally be in the point $(1,0)$.  This is just the case in our two models. 

Now let's see how the effective equation of state evolution in our models. For the power law model, the equation of state (\ref{equ:eos}) becomes
\begin{equation}
	w_e = -1 - \frac{1}{3} \frac{(4-4x-y)n}{nx-1} \,,
\end{equation}
and we also have
\begin{equation}
	w_e' =   -\frac{1+w_e }{nx-1}\bigg[ 4(n-1)x - y  -3 (n-1)x(1+w_e)  \bigg] - \frac{ny}{nx-1} \,.
\end{equation}
Thus, we have $(w_e, w_e') = (-1, 0), (-1+4n/3, 0)$ and $(-1+n, 0)$ at the points $(1,0), (0,0)$ and $(0,1)$. While  for the logarithmic model the equation of state becomes
\begin{eqnarray}
	w_e &=& -1- \frac{4-4x-y}{3(x-2)}\,, \quad \text{when} \quad \beta =\frac{1}{2} \,, \\
	w_e &=& -1 - \frac{1}{3} \frac{4-4x-y}{x-(1+\alpha)} \left( 1-\frac{\alpha }{x}\right) \,, \quad \text{when} \quad \beta =1 \,,
\end{eqnarray}
and also
\begin{eqnarray}
	w_e' &=&  \frac{1+w_e}{x-2} \bigg [ 4 x+2y  -3(1+w_e)x\bigg]  -\frac{y}{x-2}\,, \quad \text{when} \quad \beta =\frac{1}{2} \,, \\
	\nonumber
	w_e'&=& \frac{1+w_e}{x-(1+\alpha)} \left( 1-\frac{\alpha }{x}\right) ^{-1}  \bigg[ 
	(4\alpha+y )\left( 1-\frac{\alpha }{x}\right)  
	-3(1+w_e)\alpha \bigg] \\
	&&
	 + \frac{1}{3} \left[ \frac{\alpha(4-4x-y)}{x[x-(1+\alpha)]} \right]^2
	  -\frac{y}{x-(1+\alpha)} \left( 1-\frac{\alpha }{x}\right)  \,, \quad \text{when} \quad \beta =1 \,.
\end{eqnarray}
Thus, we have $(w_e, w_e') = (-1, 0), (-1/3, 0)$ and $(-1/2, 0)$ at the points $(1,0), (0,0)$ and $(0,1)$ for $\beta = 1/2$, while $(w_e, w_e') = (-1, 0)$ at $(1,0)$ for $\beta =1$.  Actually, from Eq.~(\ref{equ:eos}), one can see that  $w_e =-1$ at point $(1,0)$ no matter what forms $f(T)$ will take.

\section{Conclusions}
In conclusion, we have studied the global behavior of the $f(T)$ theory by using the nullcline method. In particular we focus on the power law model and the logarithmic model. We have found not only the equilibrium points of the system but also the initial conditions under which the system will eventually flow to the sink point. Furthermore, we find there are often bifurcation phenomena in these systems, namely the dynamical behavior of system depends on the values of parameter. It should be also noticed that we have blown up the singularity, say, $x=1/n$ by introducing a new variable. For example, in the power law model, we have defined $d\ln a/ds = (nx-1)^2$. Thus, when $x=0$, $d\ln a /ds = 0$, which means we could have a bounce solution, namely, when the expanding universe evolutes to the sate  of $x=1/n$, $H=0$ and then it will contract. Before ending this letter, we would like to emphasis that the qualitative technique of nullcline we developed in this paper is very powerful and could be used in any nonlinear dynamical system, especially in the planar system, so it deserves further studying.

It should be also noticed that the physical allowed regions of  $y ,z$ is $y, z\geq 0$, see Eq.(\ref{equ:dimless}). So, from the Friedmann equation $x+y+z=1$, we have $x\leq 1$. By definition, $x$ is only an effective Òenergy densityÓ of dark energy, see Eqs.(\ref{equ:eff}) and (\ref{equ:dimless}), so $x$ could be negative. Therefore,  the physical upper limit of $y, z$ can not be easily determined.  According to these reasons, and also to make our analysis complete, we consider larger varying regions of these variables in this paper. The bifurcation analysis studied in this paper is only to help us to know how the system evolutes differently under different regions of the model parameters.   Once these parameters are determined, e.g. by cosmological observations, one can clearly know the evolution behaviour of the system from the bifurcation analysis. For instance, it may be the right top figure in Fig.\ref{fig:fig1}. And there is no sense to talk about the bifurcation phenomena when the parameters are fixed.

\acknowledgments

CJF would like to thank Prof. Robert L. Devaney and Steven H. Strogatz for useful comments on this manuscript. This work is supported by National Science Foundation of China grant Nos.~11105091 and~11047138, ``Chen Guang" project supported by Shanghai Municipal Education Commission and Shanghai Education Development Foundation Grant No. 12CG51, National Education Foundation of China grant  No.~2009312711004, Shanghai Natural Science Foundation, China grant No.~10ZR1422000, Key Project of Chinese Ministry of Education grant, No.~211059,  and  Shanghai Special Education Foundation, No.~ssd10004, and the Program of Shanghai Normal University (DXL124).  CJF thanks the Yukawa Institute for Theoretical Physics at Kyoto University, where this work was initiated during the Long-term Workshop YITP-T-12-03 on "Gravity and Cosmology 2012".

\appendix

\end{document}